# A Hybrid Web Recommendation System based on the Improved Association Rule Mining Algorithm

Ujwala H.Wanaskar, Sheetal R.Vij, Debajyoti Mukhopadhyay

[1]Department of Computer Engineering, Padmabhooshan Vasantdata Patil Institute of Technology, Pune, India; [2]Department of Computer Engineering, Maharashtra Institute of Technology, Pune, India;[3]Department of Information Technology, Maharashtra Institute of Technology, Pune, India
Email: ujwalaw.267@gmail.com,sheetal.sh@gmail.com, debajyoti.mukhopadhyay@ gmail.com

## ABSTRACT

As the growing interest of web recommendation systems those are applied to deliver customized data for their users, we started working on this system. Generally the recommendation systems are divided into two major categories such as collaborative recommendation system and content based recommendation system. In case of collaborative recommendation systems, these try to seek out users who share same tastes that of given user as well as recommends the websites according to the liking given user. Whereas the content based recommendation systems tries to recommend web sites similar to those web sites the user has liked. In the recent research we found that the efficient technique based on association rule mining algorithm is proposed in order to solve the problem of web page recommendation. Major problem of the same is that the web pages are given equal importance. Here the importance of pages changes according to the frequency of visiting the web page as well as amount of time user spends on that page. Also recommendation of newly added web pages or the pages those are not yet visited by users are not included in the recommendation set. To overcome this problem, we have used the web usage log in the adaptive association rule based web mining where the association rules were applied to personalization. This algorithm was purely based on the Apriori data mining algorithm in order to generate the association rules. However this method also suffers from some unavoidable drawbacks. In this paper we are presenting and investigating the new approach based on weighted Association Rule Mining Algorithm and text mining. This is improved algorithm which adds semantic knowledge to the results, has more efficiency and hence gives better quality and performances as compared to existing approaches.

## 1. Introduction

With the web2.0 introduced, its use is growing up along with high speed development in infrastructure and services. Several opportunities, like sharing information and opinion with different users, came out. This did favor the event of social networks like Facebook. Nowadays, authors will share their creations with numerous readers round the globe.
Amateur-musicians will get renowned faster than ever before simply with uploading their tracks. Business world have found a lot of customers and profit within the net. The range of on-line retailers, auctions or entozoan markets can be displayed within the net. Today, each user of the World Wide Web can buy virtually any item being in any country of the planet without any place-limitations.

In fact, there's virtually endless place where we discovered a new problem within the computer universe. The number of data and things got very vast, resulting in associate degree data overload. It became an enormous drawback to seek out what the user is really probing for. Search engines partly solved that problem; but personalization of data was not considered.
Therefore, the system developers found the solution in recommender systems. Recommender systems are tools for filtering and sorting things and data. They use opinions of a community of users to assist people to determine content of interest from a doubtless overwhelming set of decisions. There's an enormous diversity of algorithms and approaches that facilitate making personalized recommendations. Two of them became very popular: collaborative filtering and content-based filtering. They are used as a base of latest recommender systems.

Appearance of mobile devices with new technologies, like GPS and 3G standards, in the market issued new challenges. Recommender systems got concerned in developing method of touristy, security and alternative areas. Trendy recommender systems are raising their recommendations accuracies by exploitation context-aware, semantic and alternative approaches. Today, recommendations are a lot of specific and personalized too. Issues of combining completely different technologies and recommending approaches for higher results can invariably exist and can be the area of interest of latest researchers.

Most of the website recommender systems that were planned earlier utilized cooperative filtering. Cooperative filtering is commonly utilized in general product recommender systems, and consists of the subsequent stages. The foremost stage in cooperative filtering is to investigate users purchase histories so as to extract user teams that have similar purchase patterns. Then suggest the products that are usually most popular within the user's cluster.

Basically opinions of community members used by the Recommender Systems (RS) in order to facilitate people to establish the knowledge possibly to be fascinating to them or pertinent to their desires. This will be achieved by drawing on user preferences and filtering the set of possible choices to a lot manageable set. Each internet Recommendation System has its own blessings and limitations.

Moreover the assignment of advocated system is to recommend things that match a user's style, so as to assist the user in selecting/purchasing things from a devastating set of selections. Such systems have huge importance in applications like e-commerce, subscription primarily based services, info filtering, internet services etc.

Recommendations are generated based on two elementary approaches. First are content based approaches in which the profiles are users and things by distinguishing their characteristic options, such as demographic information for user identification, and product information/descriptions for item profiling. The profiles are utilized by algorithms to unite user interests and item descriptions once generating recommendations [Takacs et al,]. Online page Recommendation is an energetic application space for information filtering, internet Mining and Machine Learning analysis. Another approach is cooperative recommendation that tries to seek out some users who share similar tastes with the given user and recommends websites they prefer to that user.

In this paper we are presenting a new association rule mining approach based on weighted association rule mining. We are using the weighted association rule mining algorithm in order to overcome the drawbacks of existing approaches. This approach helps user to obtain the web sites which are most relevant to them.

In the following section 2 we will discuss the different types of recommendation approaches along with their advantages and disadvantages. Section 3 presents the proposed approach for web page recommendation.

## 2. Literature Review

### 2.1. Traditional Recommendation Approaches

*2.1.1 Content-based filtering*

Content-based recommender systems work with profiles of users that are created at the beginning. A profile has information about a user and his taste which is based on how user rates the items. Generally, when creating a profile, recommender systems make a survey, to get initial information about a user in order to avoid the new-user problem. [1]

In the recommendation process, the engine compares items that were already positively rated by user with items he did not rate and looks for similarities. Those items that are mostly similar to the positively rated ones, will be recommended to the user. Content-based recommender systems mostly use tags or keywords for efficient and better filtering. In this case the profiles of other users are not essential and they don't influence the recommendations of the user, as the recommendations are based on individual information. Going in details of methods of collaborative filtering, we can distinguish most popular approaches: user-based, item-based and model-based approaches.

*2.1.2 Collaborative filtering*

Collaborative filtering became one of the most researched techniques of recommender systems since this approach was mentioned and described by Paul Resnick and Hal Varian in 1997. [2] The idea of collaborative filtering is, finding the users in a community that share appreciations [3]. If two users have same or almost same rated items in common, then they have similar tastes. Such users build a group or a so called neighborhood. A user gets recommendations to choose items that he/she has not rated before, but that were already positively rated by users in his/her neighborhood.

Collaborative filtering is widely used in e-commerce. Customers can rate books, songs, movies and then get recommendations regarding those issues in future. Moreover collaborative filtering is utilized in browsing of certain documents (e.g. documents among scientific works, articles, and magazines). [4]

Following figures 1 and 2 shows the user-based and item based methods respectively:

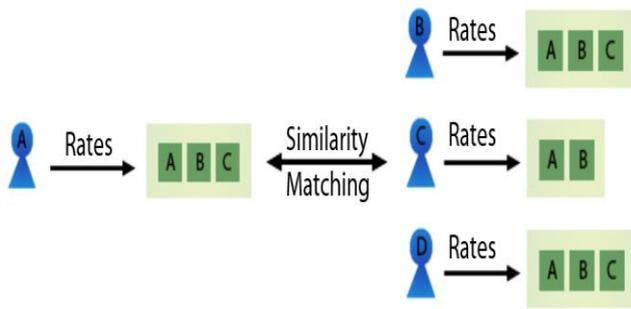

**Figure 1: User based approach**

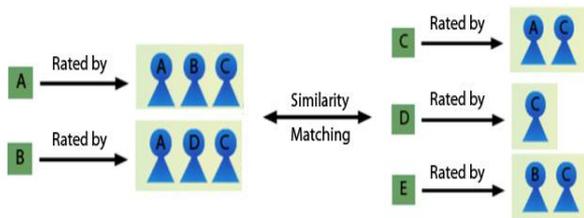

**Figure 2: Item based approach**

*2.1.3 Hybrid recommendation approaches*

For better results some recommender systems combine different techniques of collaborative approaches and content based approaches. Using hybrid approaches we can avoid some limitations and problems of pure recommender systems, like the cold-start problem. The combination of approaches can proceed in different ways:

1) Separate implementation of algorithms and joining the results.
2) Utilize some rules of content-based filtering in collaborative approach.
3) Utilize some rules of collaborative filtering in content based approach.
4) Create a unified recommender system that brings together both approaches.

Robin Burke worked out taxonomy of hybrid recommender systems categorizing them. [5]

## 2.2. Modern Recommendation Approaches

*2.2.1 Context-Aware Approaches*

Context is the information about the environment of a user and the details of situation he/she is in. Such details may play much more significant role in recommendations than ratings of items, as the ratings alone don't have detailed information about under which circumstances they were given by users. Some recommendations are more suitable to the user in evening and doesn't match his preferences in the morning at all and he/she would like to do one thing when it's cold and completely another when it's hot outside. The recommender systems that pay attention and utilize such information in giving recommendations are called context-aware recommender *systems.*

One of the biggest problems of context-aware recommender systems is obtaining context information. The information can be obtained explicitly by directly interacting with user asking him/her to fill out a form and making a survey. Although it is mostly desirable to obtain context information without making the whole rating and reviewing process complicated. Another way is gathering information implicitly using the sources like GPS, to get location, or a timestamp on transaction. [6] The last way of information extraction is analyzing users and observing their behavior or using data mining techniques.

*2.2.2 Semantic Based Approaches*

Most of the descriptions of items, users in recommender systems and the rest of the web are presented in the web in a textual form. Using tags and keywords without any semantic meanings doesn't improve the accuracy of recommendations in all cases, as some keywords may be homonyms. That is why understanding and structuring of text is a very significant part recommendation. Traditional text mining approaches that base on lexical and syntactical analysis show descriptions that can be understood by a user but not a computer or a recommender system. That was a reason of creating new text mining techniques that were based on semantic analysis. Recommender systems with such techniques are called semantic based recommender systems.

The performance of semantic recommender systems are based on knowledge base usually defined as a concept diagram (like taxonomy) or ontology.

*2.2.3 Cross-Domain Based Approaches*

Finding similar users and building an accurate neighborhood is an important part of recommending process of collaborative recommender systems. Similarities of two users are discovered based on their appreciations of items. But similar appreciations in one domain do not surely mean that in another domain valuations are similar as well.

*2.2.4 Peer-to-Peer Approaches*

The recommender systems with P2P approaches are decentralized. Each peer can relate itself to a group of other peers with same interests and get recommendations from the users of that group. Recommendations can also be given based on the history of a peer. Decentralization of recommender system can solve the scalability problem [7]

*2.2.5 Cross-lingual Approaches*

The recommender system based on cross-lingual approach lets the users receive recommendations to the items that have descriptions in languages they don't speak and understand. Yang, Chen and Wu purposed an approach for a cross lingual news group recommendation. The main idea is to map both text and keywords in different languages into a single feature space, that is to say a probability distribution over latent topics. From the descriptions of items the system parses keywords than translates them in one defined language using dictionaries. After that, using collaborative or other filtering, the system gives recommendations to users[8].

### 2.3. Challenges and Issues of Recommendation Approaches

*2.3.1 Cold-start*
It's difficult to give recommendations to new users as his profile is almost empty and he hasn't rated any items yet so his taste is unknown to the system. This is called the cold start problem. In some recommender systems this problem is solved with survey when creating a profile. Items can also have a cold-start when they are new in the system and haven't been rated before. Both of these problems can be also solved with hybrid approaches.

*2.3.2. Trust*
The voices of people with a short history may not be that relevant as the voices of those who have rich history in their profiles. The issue of trust arises towards evaluations of a certain customer. The problem could be solved by distribution of priorities to the users.

*2.3.3 Scalability*
With the growth of numbers of users and items, the system needs more resources for processing information and forming recommendations. Majority of resources is consumed with the purpose of determining users with similar tastes, and goods with similar descriptions. This problem is also solved by the combination of various types of filters and physical improvement of systems. Parts of numerous computations may also be implemented offline in order to accelerate assurance of recommendations online.

*2.3.4 Sparsity*
In online shops that have a huge amount of users and items there are almost always users that have rated just a few items. Using collaborative and other approaches recommender systems generally create neighborhoods of users using their profiles. If a user has evaluated just few items then it's pretty difficult to determine his taste and he/she could be related to the wrong neighborhood. Sparsity is the problem of lack of information. [9]

*2.3.5 Privacy*
Privacy has been the most important problem. In order to receive the most accurate and correct recommendation, the system must acquire the most amount of information possible about the user, including demographic data, and data about the location of a particular user. Naturally, the question of reliability, security and confidentiality of the given information arises. Many online shops offer effective protection of privacy of the users by utilizing specialized algorithms and programs.

## 3. Our Approach and Basics

In this paper we are trying to describe, analyze, implement and upgrade the mostly used method for web mining i.e. association rule mining [10, 11]. This technique can be easily used in recommendation systems and it is scalable [12, 13].This method gives high precision [14], and only gives binary weight to the pages that are visited i.e. to find whether the page is present or not. Usually if page is present means it is considered important. It is possible that not all the pages visited by the user are of his interest. User may visit a page but it may not have useful information for him. So factors like time spent by the user and visiting frequency of the page should be considered for the page consideration [15]. Because if user finds it interesting then only user will spend more time on it or user will visit that page frequently [16].So in association rule mining method the weight of the page is also included. This is called weighted association rule mining.

The Page weight is calculated by using following formulae [14],

$$1. Duration(p) = \frac{tot.\ duration(p)/size(p)}{\max_{Q \in T}(tot.\ duration(p)/size(p))}$$

$$2. Frequency(p) = \frac{No\ of\ visit(p)}{\sum_{Q \in T} No\ of\ visit(Q)} * \frac{1}{Indegree(p)}$$

$$3. Weight(p) = \frac{2 * frequency(p) * duration(p)}{frequency(p) + duration(p)}$$

Text Mining is a process of analyzing the text. It is useful in finding meaningful information from the given text. This technique adds semantic knowledge to the data. It is used in many applications like information retrieval, language processing, Data mining etc.There are many algorithms used for text mining. But most popular algorithm is TF-IDF.TF-IDF is term frequency and inverse document frequency. It is used to find out weighting factor i.e. how important is a word in the given document. It is used in the search engine and also to find out relevance of the documents by ranking them. This mining technique is used in the last step in the proposed system prior to generating the recommendations. By using this mining we actually filter out the pages that are highly relevant to the search query.

## 4. Proposed Algorithm

The application like web personalization is nothing but the application of machine learning techniques as well as data mining in order to build user behavior models. This application is basically used to identify the requirements of user and future interactions adapting with the main goal of increased user satisfaction.

The web recommendation systems represent unique as well as important class of the personalized web applications. The focus over the user based filtering as well as relevant information selection. Many techniques such as clustering based approaches, content based filtering, sequence and association rule based etc. are used.

In this paper we are giving hybrid recommendation approach which uses web usage mining and text mining. We are presenting the new data mining approach which is based on HITS and weighted association rule mining for the efficient web recommendation system. This method is used for providing the user a personalized web experience.

In this method, quantitative weight is assigned to each page according to amount of time the user spends on that page or the frequency of visiting that page [16]. Generally while recommending the data there is a problem of rarely visited pages or newly added pages as they would never be added to the recommendation set. So to overcome this problem in our approach we are including these pages in the data set by using HITS algorithm [16]. HITS algorithm is used to extend the data set as well as to rank the pages. Below figure 3 is showing the architecture for proposed web recommendation method.

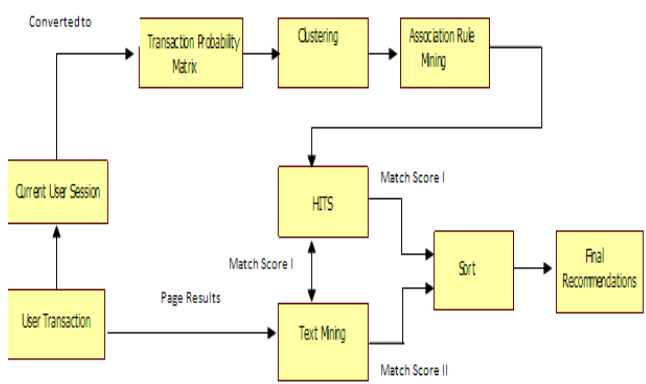

**Figure 3: Architecture of proposed approach**

As shown in above figure 3, the steps in the algorithms could be briefly summarized as follows:
Step 1: Cluster the pages based on users' usage pattern.
Step 2: Generate the seed recommendation set based on Weighted Association Rules Mining.
Step 3: Extend the seed set by clusters to generate the candidate set and apply the HITS algorithm to rank the candidate set.
Step 4: For the current user session give results in the previous step to the text mining which will generate some results. Sort the results given by step 3 and step 4 to generate final recommendation set.
The above steps are described as follows:
After the text edit has been completed, the paper is ready for the template. Duplicate the template file by using the Save As command, and use the naming convention prescribed by your journal for the name of your paper. In this newly created file, highlight all of the contents and import your prepared text file. You are now ready to style your paper.

### 4.1 *C*luster the pages based on users' usage pattern

In this algorithm, the web pages are clustered not from the content but from the pattern of their usage, assuming that users find a page very interesting and is important for the user so finds their actions [16]. In the next step we use the result of this step to extend the    recommendation set .In this step we try to cluster the pages that occur together across the sessions. Page cluster group together frequently occurring items even though they do not seem to be similar. This step forms clusters with overlapping interest of different types of users

### 4.2 Generating the Seed Recommendation Set

In this step, firstly weighted association rule of each URL is found out from web log data, these rules represents users navigation on the web. Secondly, the recommendation engine will search the top-*n* most similar weighted rules to the active user session before generating recommendation for the user. During the second part instead of exact match between the active user and rules, we use a similarity measure for finding the most similar rules.
Mining Weighted Association Rules:
Each transaction consists of the set of pages, the association rule is of the form, X->Y Where X⊂I, Y⊂I, X∩Y = Φ, Where X and Y are the two itemsets, X is the body of the rule and Y head of the rule. In this method we associate weight parameter with each page to reflect the interest of user which is called as weighted association rule mining. In this technique we just extend the traditional Apriori [17] Algorithm by adding weight as one of the parameter.

Current user session is represented as vector S of significance weight if user has accessed the page, $s_i=0$ otherwise. After this we find out the match score between association rules that generated based on navigational pattern history and current active session. This match score is calculated as follows [16],

A Hybrid Web Recommendation System based on Improved Association Rule Mining Algorithm

Dissimilarity(S,rL) = $\sum_{i:rLi>0}$ Square (2*(w(si)-w(rLi))) / (w(si)+w(rLi))

5. Match Score(S,rL) = 1-(Sqrt(Dissimilarity(S,rL) / $\sum_{i:rLi>0}$ 1)/4

6. Rec(S,X=>p) = Match Score(S,rL) * wconf(X=>p)

This recommendation system is an online component of personalization system which determines which items to be recommended to the user. The recommendation score is calculated by multiplying match score and weighted confidence. Finally top n-most similar pages are sorted to be used in the next phase.

### 4.3 Extending the Seed Set and Apply Hits

Generally problem with the recommendation system is that its recommendation accuracy decreases as dataset increases also rarely visited pages or newly added pages are not included in the recommendation set. These pages should be included in the recommendation set otherwise they would never get recommended.

To overcome this problem we have used seed recommendation set generated in the previous step as input to this step. We extend this set to generate candidate set. For each page in the seed set candidate is supplemented with the pages that are in the same cluster. A graph is generated from the pages included in the candidate set by connecting them with the link that exists, results in the connectivity graph which represents improved navigational pattern. This process of obtaining the connectivity graph is same as that used by HITS algorithm [18] to find Authority and Hubs. So we take the advantage of HITS algorithm to identify hubs and authority pages within that clusters which allows us to rank pages within the clusters. Here only Hub measures are considered as it may link to many authority pages [19]. Using this hub value we will rank the candidate set pages in online module to form Match Score I.

### 4.4 Apply Text Mining On Results and Generate Final Recommendation

In this step, the text mining is done for more approximated results. The TF-IDF algorithm is used for this. The results of the above step as well as the page results for the current user session are given as the input to this stage. Now the results given by this stage and previous step are sorted to generate final recommendation set.

### 4.5 HITS Mathematical Model

According to Jon Kleinberg's a higher $h_i$ number as being better hubs. Given the weights $\{a_i\}$ and $\{h_i\}$ of all the nodes in $S_Q$, we dynamically update the weights as shown in the following figure 4:

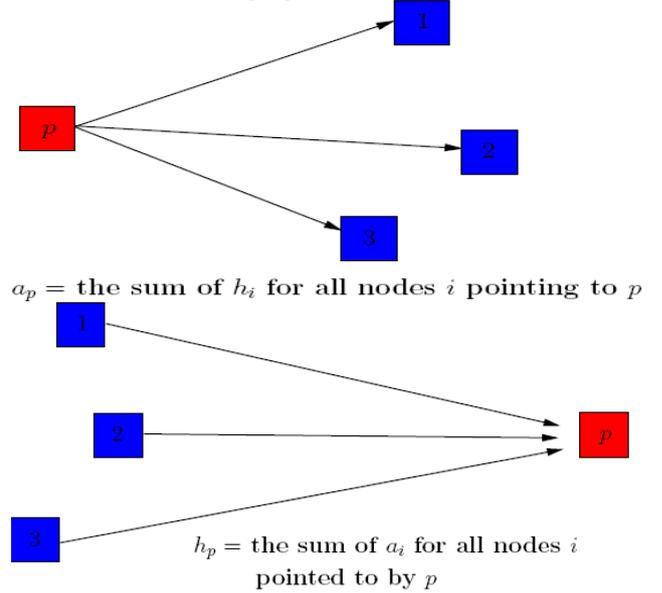

$a_p$ = the sum of $h_i$ for all nodes $i$ pointing to $p$

$h_p$ = the sum of $a_i$ for all nodes $i$ pointed to by $p$

**Figure 4: Dynamic weight updation**

A good hub increases authority weight of pages it points. A good authority increases the hub weight of the pages that point to it. The idea is then to apply the two operations above alternatively until equilibrium values for the hub and authority weights are reached. Let A be the adjacency matrix of graph SQ and denote the authority weight vector by v and the hub weight vector by u, where

$$v = \begin{bmatrix} a_1 \\ a_2 \\ a_3 \\ a_4 \end{bmatrix} \quad \text{and} \quad u = \begin{bmatrix} h_1 \\ h_2 \\ h_3 \\ h_4 \end{bmatrix}$$

Let us notice that the two update operations described in

$$\begin{cases} v = A^t \cdot u \\ u = A \cdot v \end{cases}$$

the pictures translate to:

If we consider that initial weights of the nodes are

$$u_0 = \begin{bmatrix} 1 \\ 1 \\ \vdots \\ 1 \end{bmatrix} \quad \text{and} \quad v_0 = A^t \cdot \begin{bmatrix} 1 \\ 1 \\ \vdots \\ 1 \end{bmatrix}$$

Then, after *k* steps we get the system:

$$\begin{cases} v_k = (A^t \cdot A) \cdot v_{k-1} \\ u_k = (A \cdot A^t) \cdot u_{k-1} \end{cases}$$

**Algorithm 1:** Under the assumptions that $AA^t$ and $A^tA$ are primitive matrices, following statements hold:

i. If $v_1, ..., v_k$ is the sequence of authority weights we have computed, then $V_1, ..., V_k$ converges to the unique probabilistic vector corresponding to the dominant eigen value of the matrix $A^tA$. With a slight abuse of notation, we denoted in here by $V_k$ the vector $v_k$ normalized so that the sum of its entries is 1.

ii. Likewise, if $u_1, ..., u_k$ are the hub weights that we have iteratively computed, then $U_1, ..., U$ converges to the unique probabilistic vector corresponding to the dominant eigen value of the matrix $AA^t$. We use the same notation, that $U_k = (1/c)u_k$, where $c$ is the scalar equal to the sum of the entries of the vector $u_k$.

So authority weight vector is the probabilistic eigenvector corresponding to the largest eigenvalue of $A^tA$, while hub weights of the nodes are given by the probabilistic eigenvector of the largest eigenvalue of $AA^t$:

**Algorithm:**

1. The matrices $AA^t$ and $A^tA$ are real and symmetric, so they have only real eigenvalues.

2. **Perron Frobenius**. If $M$ is a primitive matrix, then:
   i. The largest eigen value $\lambda$ of $M$ is positive and of multiplicity 1.
   ii. Every other eigen value of $M$ is in modulus strictly less than $\lambda$
   iii. The largest eigen value $\lambda$ has a corresponding eigen vector with all entries positive.

3. Let $M$ be a non-negative symmetric and primitive matrix and $v$ be the largest eigenvector of $M$, with sum of its entries equal to 1. Let $z$ be the column vector with all entries non-negative, then, if we normalize the vectors $z, Mz,...,M^kz$, then the sequence converges to $v$.

HITS algorithm is in the same spirit as PageRank. They both make use of the link structure of the Web graph in order to decide the relevance of the pages. The difference is that unlike the PageRank algorithm, HITS only operates on a small subgraph (the seed $S_Q$) from the web graph. This subgraph is query dependent; whenever we search with a different query phrase, the seed changes as well. HITS rank the seed nodes according to their authority and hub weights. The highest ranking pages are displayed to the user by the query engine.

## 5. Experimental Analysis

To evaluate the effectiveness of the method, performance is measured using two factors like precision and coverage [18].Recommendation precision means number of correct recommendations i.e. proportion of relevant recommendations to the total number of recommendations. Precision is given by the formula,

Precision = (T (p) ∩R (p))/R (p)

Coverage of the system is the proportion of relevant recommendations to the all pages that should be recommended. Where R (p) is recommendation set and T (p) is session. Precision of the recommendations are measured for varying number of recommended pages.So based on above proposed system we have worked on practical evaluation using the JAVA, J2EE. We have done implementation through the web application as shown in the following figure 5:

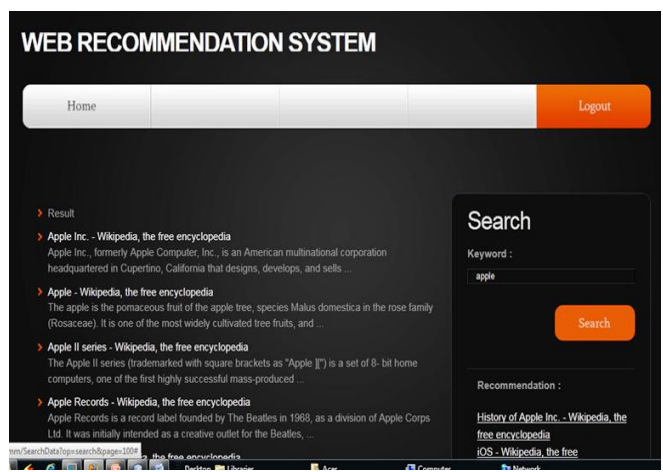

**Figure 5: Search Results of web page Recommendation system.**

Following graphs are showing the performance evaluation of the existing algorithm i.e. based on only weighted association rule mining and our proposed approach based on the weighted association rule mining and text mining .The following table readings and graphs show the improved performance of as compared to existing cases.

Comparative Study of precision rate between proposed and existing method based on number of pages ranked.

Table 1 and 2 shows the readings we got during our practical analysis and figure 6 and 7 shows the graph for those readings:

The table and the following graph shows that the proposed approach shows improved results as compared to the previous approach. So approach is efficient as compared to the previous existing approaches.

**Table 1: Precision comparison readings**

| No of Recommended Pages | Precision | |
|---|---|---|
| | Proposed Algorithm | Previous Algorithm |
| 1 | 100 | 100 |
| 2 | 50 | 33.33 |
| 3 | 100 | 100 |
| 4 | 100 | 87.5 |
| 5 | 80 | 60 |
| 6 | 58.33 | 50 |
| 7 | 57.14 | 50 |
| 8 | 85.71 | 62.5 |
| 9 | 33.33 | 27.77 |

**Table 2: Coverage comparison readings**

| No of Recommended Pages | Coverage | |
|---|---|---|
| | Proposed Algorithm | Previous Algorithm |
| 1 | 50 | 50 |
| 2 | 66.66 | 66.66 |
| 3 | 65.5 | 16.66 |
| 4 | 100 | 50 |
| 5 | 100 | 83.33 |
| 6 | 57.14 | 50 |
| 7 | 100 | 83.33 |
| 8 | 62.5 | 16.66 |
| 9 | 33.33 | 20 |
| 13 | 62.5 | 38.46 |

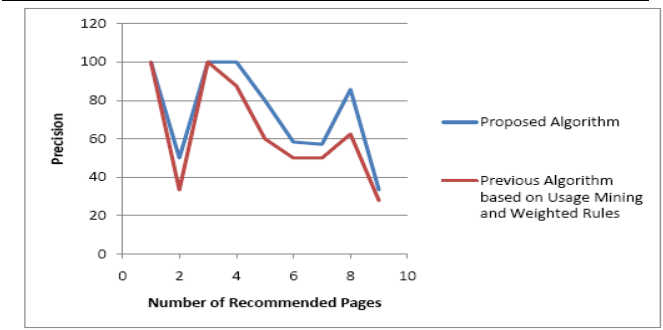

**Figure 6: Precision comparative Analysis**

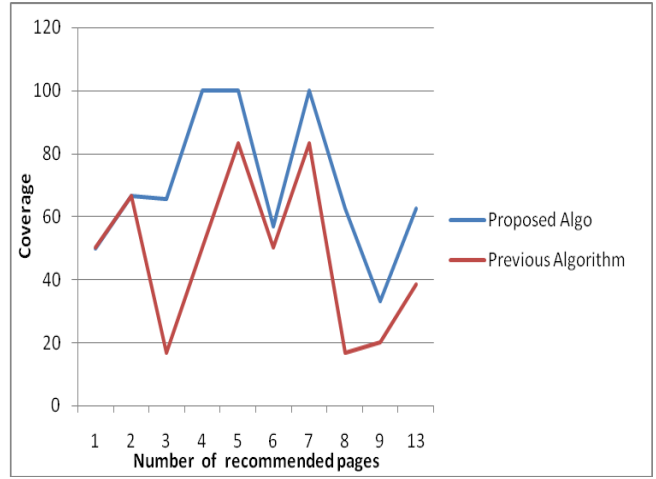

**Figure 7: Coverage Comparative Analysis**

## 6. Conclusion and Future Scope

This paper proposes as new web recommendation system based on weighted association rule mining and text mining. In this approach, weight is assigned to each page to show its importance depending on the time spent by each user on a particular page or visiting frequency of each page. To add semantic knowledge to the data to be recommended, text mining is used.

One of the challenging problem with recommendation system is that pages that are newly added or rarely visited. They are generally not included in the recommendation set, so in this approach we have added these pages to the recommendation page set. The performance of the system is evaluated under different settings and in comparison with the previous method which is based only on the weighted association rule mining

Web recommendation system is used to recommend pages to users. The application can be used for personalized recommendation to give personalized recommendation based on users browsing history. This system can be used in search engines to give recommendation to users based on the users search keyword so that system will make proper recommendation, filtering unrelated information. Recommendations can be given for individual sites/ports or generalized sites.

Throughout this paper we have discussed many aspects of research for web recommendation systems. We have presented related work, problems associated with existing methods as well as literature study over various research methods in the same domain. Based on existing limitations, in this paper new mining approach based on combination of weighted association rule mining and text mining is presented which is showing the better performance improvement as compared to the existing methods. For the work we suggest to apply this method under cloud computing environment.